# A novel approach for protection of accounts' names against hackers combining cluster analysis and chaotic theory

Desislav Andreev
*Dep. Computer Science,
Faculty of Computer Systems
and Technologies,
Technical University of Sofia*
desislav.andreev@gmail.com

Simona Petrakieva
*Dep. Theory of Electrical
Engineering,
Faculty of Automation
Technical University of Sofia*
petrakievas-te@tu-sofia.bg

Ina Taralova
*Laboratoire des Sciences du
Numérique de Nantes
LS2N  UMR 6004
Ecole Centrale de Nantes*
ina.taralova@ec-nantes.fr

**Abstract**

*The last years of the 20$^{th}$ century and the beginning of the 21$^{th}$ mark the facilitation trend of our real life due to the big development and progress of the computers and other intelligent devices. Algorithms based on artificial intelligence are basically a part of the software. The transmitted information by Internet or LAN arises continuously and it is expected that the protection of the data has been ensured.*

*The aim of the present paper is to reveal false names of users' accounts as a result of hackers' attacks. The probability a given account to be either false or actual is calculated using a novel approach combining machine learning analysis (especially clusters' analysis) with chaos theory. The suspected account will be used as a pattern and by classification techniques clusters will be formed with a respective probability this name to be false.*

*This investigation puts two main purposes: First, to determine if there exists a trend of appearance of the similar usernames, which arises during the creation of new accounts. Second, to detect the false usernames and to discriminate those from the real ones, independently of that if two types of accounts are generated with the same speed.*

*These security systems are applied in different areas, where the security of the data in users' accounts is strictly required. For example, they can be used in on-line voting for balloting, in studying the social opinion by inquiries, in protection of the information in different user accounts of given system etc.*

## 1. Introduction

Nowadays the most expensive thing is the information itself. Some systems contain confidential information that should be kept safe and secure from hackers' attacks at all time. The consequences of unauthorized access can be devastating from identity theft problems to loss of irreplaceable data. When some data is changed or deleted on purpose or by chance, it creates chaos, calling into question about the reliability or accuracy of all data in the system. This leads to delay in the terms for the realization of the important projects in companies, which directly causes financial losses. Usually, users' accounts are secured with encryption of their username and password, but this is not absolutely reliable and sometimes there has to be an additional verification about the account – false or actual. In general, Windows has a build-in User Account Control which helps to the respective organization to configure its security policies and guarantee the safety of self-data [1]. Some software companies like Oracle [2], Microsoft [3], Apple [4] and etc., have developed their own software for observation and detection of false usernames and passwords. Based on the reasons, mentioned above, to increase the reliability and safety of the data in users' accounts in the present paper a new approach to reveal false users' accounts is proposed. It combines the cluster analysis technique and the chaos theory for recognition of each account as either actual or false.

Based on this the approach uses clustering for grouping and placing possible false user accounts as obvious outliers. The clustering is based on the concepts of similarity and distance. During this process clusters that consist of individuals who have common features with each other are generated. The similarities between the inputs are estimated by a so-called distance function. An important basic feature of clustering is that the groups are not given a priori, but they form during the evaluation process using the groups found until the current moment of the analysis.

The probability in each cluster for a given account to be actual or false **is** determined using cluster analysis and chaos theory.

Chaotic systems are characterized by their extreme sensitivity to small changes in the initial conditions and/or parameters, which makes them look like any random function [5, 6]. However, the deterministic feature of the chaotic functions is characterized by their repetitively, i.e. identical initial conditions and parameters would generate identical chaotic sequences (when run on the same processor, if the system is highly chaotic). In more recent papers [7, 8, 10], original chaotic maps have been designed and proved to be efficient chaotic pseudo-random number generators: (CPRNG): each different tuning of the initial conditions or parameters gives rise to a different chaotic sequence (or carrier), and therefore, to a different pseudo-random key. In the current paper, a chaotic pseudo-random number generator is used to simulate efficiently different accounts with false names.

A direct impact and applicability of the suggested approach can be illustrated in revealing the false names of accounts, which for instance could lead to increase the voters' participation, since voters will be able to securely use all modern tools for communication (smartphones, tablets, netbooks, etc.) to vote even when they will be far from their home cities (during bank holidays for instance), and also, simultaneously decrease the election costs usually quite high in traditional paper voting. Besides this can facilitate citizen rights for voting for disabled people who could vote from their homes even when they live far away from the voting centers.

The present paper is organized as follows. In the next Section 2 the main requirements for chaotic generator design are proposed. Section 3 describes the proposed cluster analysis method for determining the probability a considered user's account to be either false or actual. The applicability of suggested approach, combining clusters analysis chaotic generator is illustrated on example in Section 4. The paper finished with conclusion remarks about advantages and disadvantages of the new proposed security checking approach as well as the most general areas of its applicability.

## 2. Chaotic generator to create accounts

Chaotic systems are attracting attention due to their strong sensitivity to small variations in the initial conditions and parameters (known as the "butterfly effect"). There are very good candidate whenever security is required, since they can exhibit excellent statistical (i.e. "random") features, while being quite simple for analytical design. Indeed, complex (e.g. chaotic) behavior can be exhibited also by some apparently simple piece-wise linear systems. At the same time, these systems [5, 6] could be considered – and analyzed - as switched systems from control theory point of view. So, switched systems in chaotic regime appear to be perfect candidates able to generate rapidly different chaotic pseudo-random sequences: in principle, it is enough to change slightly a parameter or initial condition to obtain a different chaotic bitstream.

Nonlinear chaotic functions have already been proposed [7, 8, 9, 10, 11, 12] as interesting alternative solutions to improve security, since small deviations from the initial conditions - or the parameters - may lead to exponential divergence of the corresponding (generated) trajectories. Since, complex (e.g. chaotic) behavior can be exhibited even by some piece-wise linear (PWL) maps [5, 6, 9, 10, 12]. It makes them particularly attractive for different real life applications such as Secure Communications, Information Encryption or Secure Electronic Transactions. The above maps, when used in chaotic regime, show up as perfect candidates to generate uncorrelated (independent) output signals. Under some conditions, the latter can be considered as pseudo-random sequences, and applied as chaotic pseudo-random generators or carriers for secure information transmission. The main difficulty arises from the fact that the majority of chaotic maps that are widely spread in the literature (logistic map, tent map etc.) are not naturally suitable for CPRNG design, and most of them don't exhibit satisfactory randomness properties - mainly because of their weak chaoticity and/or collapsing effect when numerically simulated. However, a judicious coupling of tent, and logistic maps has been proposed recently for random number generation purposes [9].

Here many practical problems arise, from the choice of the structure and parameters of the particular chaotic generator, and its parameters, to the best coupling while satisfying predefined criteria for randomness and security. A comparative analysis of the chaotic functions is to be carried out to select the most suitable ones according to predefined criteria from system theory point of view: uniform probability density function; chaotic attractor and basin of attraction; positive Lyapunov exponents; correlation function analysis.

The selected chaotic functions must be tested for their capability to be (successfully) used as pseudo-random numbers generators to generate independent sequences. As for the application, the sequences have to be binarized, and verified for randomness, i.e. that neither future samples can be predicted from the current ones, nor past samples can be retrieved. However, the results depend on the way that the binarization has been performed, and are highly sensitive to it.

For that reason, besides the tests for randomness, correlation analysis needs to be carried out to prove that past and future bits are indeed uncorrelated. In addition the quasi-random bitstream should keep its pseudo-random properties for small parameter modification, i.e. the chaotic behaviour has to be preserved.

The task therefore for this part of the study has been to select robust chaotic generator that could be used successfully as a pseudo-random number generator, while satisfying the above mentioned properties.

The particular interest in chaotic systems is due to the fact that they are complex and exhibit at the same time specific features, since the complexity could be attributed to different sources (combined nonlinear dynamics; specific nonlinearities etc.), therefore specific methods are to be used for the design and evaluation of their performances. Some of these tools concern the properties of the chaotic attractor in the phase plane, when used as CPRNG: the chaotic attractor has to be dense, without holes, with uniform probability distribution in the whole phase space, which can be investigated for instance using Markov matrix of transitions.

Another tool is the bifurcation diagrams analysis which shows the evolution of the system dynamic behavior with changes in the parameters, in other words, their impact on the complex behavior of the system, in particular for which parameters chaotic behavior can be observed or not. This analysis is very important to choose the appropriate parameter tuning. The main concept is that of a robust chaos, which means that the chaotic properties have to be preserved when the parameters have been changed, within the same parameter "window".

There are different ways to improve the properties of the chaotic map when used as CPRNG. Some of the directions are to increase the order of the chaotic map; to apply different couplings of individual chaotic 1D maps, or to perform chaotic undersampling ([8]).

From signal processing point of view, a detailed correlation analysis should be performed in addition to demonstrate that each sample of the chaotic sequence is correlated with itself, but not with the others and the correlation strongly decreases with time (short memory). Last but not the least, pseudo-random features of the CPRNG have to be validated by NIST (National Institute of Standards and Technology) tests for randomness.

Following the previous requirements, a ring mapping $M_p$ of $p$ tent maps proposed in [8], has been selected as efficient CPRNG, where the state variables $x_{n+1}^j$ are defined at discrete time "$n+1$" and the index $j$ defines the $j^{th}$ state component.

$$M_p : \begin{aligned} x_{n+1}^1 &= 1 - 2|x_n^1| + k^1 \times x_n^2 \\ x_{n+1}^2 &= 1 - 2|x_n^2| + k^2 \times x_n^3 \\ &\vdots \\ x_{n+1}^p &= 1 - 2|x_n^p| + k^p \times x_n^{p+1} \end{aligned}$$

where the coupling coefficients $k^i \in [-1, 1]$.

Escaping (diverging) trajectories are fed back to the torus $J^p = [-1, 1]^p \subset \Re^p$ by

$$\begin{aligned} &if\ x_{n+1}^j > 1\ \text{substract}\ 2 \\ &if\ x_{n+1}^j < -1\ \text{add}\ 2 \end{aligned}, \quad j = \overline{1, p}$$

In this way, the trajectories are bounded to the torus, where they are constrained to evolve.

This mapping allows to generate parallel bitsreams, where each state component is independent from the others.

## 3. Basis of Cluster Analysis

The main idea of cluster analysis consists of finding similarities between the considered features of real objects, words, expressions, etc. This is a very important concept nowadays. The technique can be applied for investigation in the continuous time-series cases like the discrete sequences and also the discrete event systems [13]. The physical nature of the objects and the type of their mathematical model (discrete or continuous) are not significantly different. Essentially, the cluster analysis is a technique for analysis and forming of separate groups of objects with similar features, thus creating occasionally an observable pattern.

Often the analyzed object becomes more complex and - as we said already, when it comes in bigger numbers, the more complicated and nontrivial analysis is required for capturing a certain pattern. If we can convert successfully the object into a discrete forming representation sub-sequences, then we can assume that each sub-sequence is a pattern, representing a part of the whole input sequence. Then we can analyze and prepare a set of routines that match a given discrete input instance to an already stored one. Comparing both requires a certain metric that describes how they are similar to each other. Knowing this we can manipulate the inputs later to get them as close as possible basically making a conclusion about a detected recognizable known pattern.

Nowadays, there exists a variety of different clustering methods for forming separate clusters. In our case – to detect false usernames for the respective user's account – the most appropriate for grouping the data in datasets are three of them: **k-means clustering, mini-batch k-means and agglomerative clustering** [13].

**K-mean clustering** is probably the most popular, because it is simple for implementation and provides good quality output for low-dimensional input data. The main and the most popular approach of this procedure is called the "Lloyd's algorithm": Partition $n$ number of observations in $k$ groups (clusters), where each observation has to belong to only one cluster with the nearest mean, serving as a prototype

(centroid) of that cluster. In other words, the distance function of the considered observation has to be the minimum one with respect to the cluster, where it belongs. The main disadvantage of this method is that the computational time is too long when the datasets are too big and multidimensional.

**Mini-batch k-means clustering** is an improved version of the k-means method. Here the computation time is reduced, but the provided output is almost the same as with the standard k-means evaluation. But in a contrast, here the update of the centroids is being done on a per-sample basis.

**Agglomerative clustering** uses a bottom-up approach, where every data (observation) in the dataset is considered as a separate cluster in the beginning. Pairs of clusters are being merged as one moves up the hierarchy. It's easy to see, that this method requires a lot of evaluation time, since it deals with every observation on the start and also does a greedy strategy for computation.

The main difference between these clustering methods is the computational time and not many deviations would be observed during the experiments.

In this paper the main aim is to find the best suited method for clustering certain data – reveal the false names of user accounts among all the usernames in dataset. So the three methods mentioned above will be applied for clustering certain data. But each of them can be used for clustering uncertain data, having in mind that in most cases the data in real world is raw and unedited.

To estimate the workability of these methods and to compute the quality of the clustering and in order to interpret the consistency of the cluster data we will use the **Silhouette method**. There, the assessment of the quality of the clustering methods described above is made by the following quantitative numerical indicator:

$$s(i) = \frac{b(i) - a(i)}{\max\{a(i), b(i)\}} \quad (1)$$

where: $a(i)$ is the average dissimilarity of $i$-th observation in a considered dataset with all other data in the same cluster.
$b(i)$ is the lowest average dissimilarity of $i$ to any other cluster, of which $i$ is not a member.

As a conclusion based on the **Silhouette** assessment indicator it is clear that having a small number for $a(i)$ means a bigger $s(i)$, so the data is well matched. If $s(i)$ is close to negative one, then $i$ would better be placed in a neighboring cluster. If the result is zero, then $i$ is on the border of two clusters.

An interesting approach for detecting fake content was presented in [15], where the text is being randomly taken from different sources and concatenated in a new and realistically looking article or document. In the paper are proposed two methods for detecting fake content. The first is based on the classical rules in the language models but the second one improves short range information between the words using relative entropy scoring. They are validated using the domain independent model based on Google's 3- and 4-grams models. First method gives fairly good results, but the second one improve results with respect to the words' generators as a Markovian text generators. This approach can be considered as a bridge between plagiarism and stylistics models for detection of the face text content.

Redesigning the techniques, suggested in [15], at least for account *usernames* provides the following two basic ideas:

- If the *usernames* should be created only by the real names of the users - for example an online voting website, than randomly generated names could appear if software techniques for fake content generation, mentioned in [15] are applied. Such a behavior should be detected easily, because the autonomous string creation will lead to unexpected same-letter multiplication, for example: *johnssssmith* instead of *johnsmith*[1]

- This is obviously not enough for detecting false names, because the administrator or the MIS should not limit that much the naming convention of the users. Moreover if the fake content detector, based on relative entropy in [15] is able to mark suspicious names, this will not help in the detection of the so-called *gibberish* (also-known-as random text), for example: *fsdfasdf* instead of *fionaalder*[1]

Having these two points analyzed it becomes clear that certain algorithm for detection and representation of possibly-false names is required. With the machine learning techniques of clustering strings and simple probability measuring, the outliers in a given array of names – possibly a database on a web-server, will be detected and marked as possible intruders. The continuously arising number of the outliers in time is an obvious trend, which must be a signal for high security risk. Defining these requirements provides the opportunity for continuing this analysis.

The clustering technique is based on the concepts of similarity and distance, while proximity is determined by a distance function. It allows the generation of clusters where each of these groups consists of individuals who have common features with each other. An important point in this technique is that the groups are hardly defined a priori and the usage of the algorithm has to be verified by the person, using the system. One of the best choices for a clustering algorithm is the **k-means** one, but it's required that the user sets the number of the expected patterns. It provides a quick evaluation with good clustering quality and its implementation is easy to support [13]. In Section 2 we defined that with the methods of the chaos analysis we could define a number of suspected account patterns, e.g. the number of clusters we required as an input for the **k-means** algorithm. Since this methodology is currently not analyzed throughout, a simplified version of the cluster analysis and the false name probability will be used.

---

[1]The usernames are randomly created and only suggested for presentation reasons, so they do not represent a 100 % example of user name creation in different online services.

Since the current application is just a proof-of-concept we could limit ourselves in using the traditional technique, although in future systems an automated approach could become a better solution. Let's say we discover more patterns on the run - in that case we would require re-evaluation of the algorithms with updated number of *k*-s. A study providing solid results is already conducted and we see that such approach is meaningful, but will not provide any asset to the current research [14].

As we discuss in Section 2 it is obvious, that the chaos analysis is not always applicable. This brings us to the alternative - to ignore the step of a priori cluster number definition and proceed with a technique, which gets the output to similar as the **k-means** one. We already discussed this above – the **agglomerative clustering**. It is obvious that the evaluation of it will take $O(n^3)$ time complexity and around $O(n^2)$ memory space, because of its nature – the *agglomerative* strategy branch of the hierarchical clustering methods [16]. In a contrast, the **k-means** approach is placed into the *divisive* strategy branch. The initial number of clusters will not be defined directly from a previous procedure, as it was for the **k-means**, but will be initialized with the number of observations – each observation is one cluster in the beginning. Since the procedure is recursive, we have to define a logical end of the function, said otherwise – to provide the number of clusters that have to be formed in the end. In order to not define it heuristically, during the comparison task we will use the same number as it is provided for **k.** If it is not provided, then the final number of clusters will be two. For the area of our research this will do fine, but for big systems there has to be an optimization or a usage of different, most probably a heuristic, algorithm. The major optimizations for the clustering strategies are based on their metric solution for getting the distance between to inputs and the distinct advantage of the hierarchical clustering strategies is exactly the usage of any valid distance measure, whatsoever [17]. A second characteristic to check is the linkage criterion, which is key for the hierarchical clustering's quality [18]. The reason being so is the fact that such kind of clustering approaches creates hierarchy of clusters (where the name comes from) and the strategy, on which the input groups are linked with each other, is crucial for the future maintenance of them. Some of the most famous strategies are:

- **Average linkage** – It minimizes the average of the distances between all observations of pairs of clusters.
- **Maximum or complete linkage** – It minimizes the maximum distance between observations of pairs of clusters.
- **Ward's minimum variance method** – We can base our comparison analysis exactly on this strategy, because it minimizes the sum of the squared differences within all clusters. It is a variance - minimizing approach and in this sense is similar to the k-means objective function.

Defining these points our overall algorithm could be observed in two approaches that differ on their clustering technique and thus providing the opportunity for simultaneous comparison.

## 4. Illustrative example how to reveal false users' accounts

The purposes of the current example are the following:

a) To determine if there exists a trend of generation of false accounts, that arises during the creation of new accounts.

b) To detect the false accounts and to discriminate those from the real ones independently of that if two types of accounts are generated simultaneously.

Following the points from above the algorithm at first will check the probability of a *name* to be random generated (written with the English alphabet). This could be evaluated easily with some basic calculation of the relationship between the number of vowels and consonants in a given input. With simple observations over the names in Bulgaria, we could simply parameterize as follows. The unique letters are 35 – 45 % in a name. The names are averagely 10-15 letters long. The vowels are between 45 and 55 percent in average. This will be used as the first dimension of our future clusters.

In addition to these probabilities, we will assign a possible number of close names to a given one, using a similarity measurement algorithm [16].

For our purpose in the present paper we will use only two algorithms for detecting the fake usernames of accounts – Jaro's algorithm and *Jaro-Winkler's* algorithm.

By *Jaro* algorithm the distance between two compared strings $s_1$ and $s_2$ are determined by the following formulae:

$$sim_{jaro}(s_1, s_2) = \frac{1}{3} \cdot \left( \frac{c}{|s_1|} + \frac{c}{|s_2|} + \frac{c-t}{c} \right) \quad (2)$$

where: *sim* is the similarity function between two given strings $s_1$ and $s_2$, measured firstly with *Jaro*'s formulae;
c is the number of common (matching) characters;
t is the minimum number of single-character transpositions required to change one string to another.

The *Jaro*'s formula (2) is used in computer science and mathematics to measure the distance between two text strings or sequences - $S_1$ and $S_2$. Each character of string $S_1$ is compared with all its matching characters in $S_2$. A possible example for the values of *c* and *t* could be observed below:

| $S_1$ | P | A | U | L |
|---|---|---|---|---|
| $S_2$ | P | U | A | L |

Table 1

where: $c = 4$ and $t = \frac{2}{2} = 1$

The complexities of the Jaro's algorithm are $O(|S_1| + |S_2|)$.

One improvement of *Jaro*'s algorithm for calculation the distances between two strings *Jaro*'s is so called *Jaro-Winkler* algorithm. The latter is a good approach which is based on the following formulae:

$$sim_{wink}(s_1, s_2) =$$
$$= \begin{cases} sim_{jaro}(s_1, s_2), \text{ when } sim_{jaro}(s_1, s_2) < 0.7 \\ sim_{jaro}(s_1, s_2) + \frac{l}{10} \cdot (1 - sim_{jaro}(s_1, s_2)), \text{otherwise} \end{cases} \quad (3)$$

where the new parameter *l* is the number of agreeing characters in the length of common prefix at the start of the string. Its length can reach maximum of characters. This calls for the fact that usually the errors occurs in the prefixes in the names.

It is obvious that *Winkler*'s formulae (3) is based on the *sim* function from *Jaro*'s formula. *Winkler*'s will produce more favourable ratings for names that match from the beginning (defined by the additional prefix calculation in the second branch of (3)).

For our purpose of cluster analysis, each name will have the number of the similar to it measurements, and this will be the second dimension of our cluster. For the clustering we will use the essence of Lloyd's algorithm, which is a typical **member of k-mean algorithms**. It was construct from Lloyd and noted in a document in 1957, although this documents were published not until 1982 [19]. The main aim of this algorithm is to divide *n* number of in initial data points (for example *n* strings) in *k*-number of clusters. Each data point has *d* dimensions, i.e. it considers $R^d$ *s*pace. The criterion for belonging of the given data point to the associative cluster is to minimize the mean squared distance between this data point and the nearest center of the cluster where it belongs.

In this paper we apply the idea of the proposed in [20] simple and efficient implementation of Lloyd's algorithm, so called filtering algorithm. The latter requires a *kd*-tree as the only major data structure. The practical efficiency of this algorithm establishes in the fact that it runs faster as the separation between clusters increases.

The essence of this algorithm is following:

$$C_k = \{x_n : \|x_n - \mu_k\| \leq all \ \|x_n - \mu_l\|\} \quad (4)$$

$$\mu_k = \frac{1}{C_k} \sum_{x_n \in C_k} x_n \quad (5)$$

where: $\mu_k$ and $\mu_l$ are given sets of centroids;
$C_k$ is $k^{-th}$ cluster;
$x_n$ is $n^{-th}$ measurement.

In parallel to the **k-means** approach we will execute our algorithm's steps switching the clustering evaluation to **agglomerative clustering.** As discussed in [21] and above in Section 3 the Ward's method is recommended. The latter bases on the bottom up approach of Lance-Williams algorithm. The minimum variance criterion $d_{ij}$ again is based on the Euclidean distance and it could be presented as:

$$d_{ij} = d(\{X_i\}, \{X_j\}) = \|X_i - X_j\|^2 \quad (6)$$

Since it is a Lance-Williams algorithm, (6) must be transformed to a recursive representation of the distance $d_{(ij)k}$ between the new cluster $C_i \cup C_j$ and $C_k$ according to the following formulae:

$$d_{(ij)k} = \alpha_i d_{ik} + \alpha_j d_{jk} + \beta d_{ij} + \gamma |d_{ik} - d_{jk}| \quad (7)$$

where: $d_{ij}$, $d_{ik}$, $d_{jk}$ are the pair-wise distances between clusters $C_i$, $C_j$, $C_k$, respectively. $C_i$, $C_j$, $C_k$ are the $i^{-th}$, $j^{-th}$ and $k^{-th}$ clusters respectively.

$\alpha_i$, $\alpha_j$, $\beta$, $\gamma$ are parameters that depend on the clusters' sizes and they determines as follows:

$$\alpha_l = \frac{n_l + n_k}{n_i + n_j + n_k} \quad (8.1)$$

$$\beta = \frac{-n_k}{n_i + n_j + n_k} \quad (8.2)$$

$$\gamma = 0 \quad (8.3)$$

The explanation of the reason which make assumption (8.3) is given in [21]. The clusters are updated to contain the points that are closest in distance to each centroid. For every set of clusters the centroids are recalculated as the means of all points belonging to a cluster.

For the initialization we will use lists of existing names, where the number of the names will be the **k** parameter of the clustering algorithm for the **k-means** and the final cluster count for **agglomerative**. In addition to these names we will feed into the algorithm a list of random-name variations for each input, where a chaotic random generator from Section 2 could be used. The parameter with the highest probability to be a real name automatically will be assigned as a centroid for each cluster and its variations will be scattered around it. One cluster would have the values with the highest probability not to be a real name or such that is not provided with the training data. This will mark the inputs we have to check immediately. Nevertheless, the main idea for this algorithm is to mark suspicious input that will be double-checked by the system's administrator.

In the current research the clustering method **k-means** - from machine learning theory is applied in order to partition **n** data or measurements into **k** clusters (number of Voronoi cells). The idea is that each measurement (or observed data) is attributed to a corresponding **k** – class, in according to some quantifiers, as the nearest mean. In the considered case, **n** is 200 and **k** is 3. The evaluations results are shown in Figure 1.

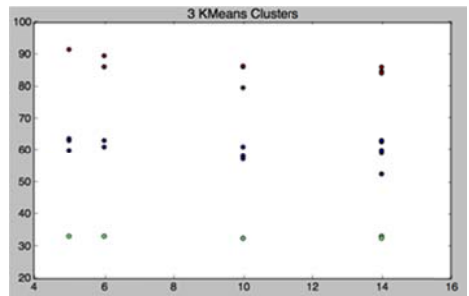

Figure 1

Let's conduct the experiment for the other clustering approach - **agglomerative.** We will partition the same data and in the end we should reach the same **k** as in the execution above, although with chaos analysis we could expect different output. The results of this analysis are presented in Figure 2:

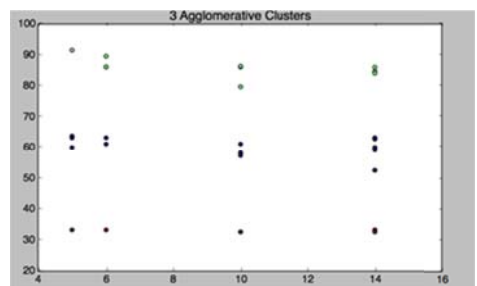

Figure 2

On the X-axis of the figure, we can observe the number of similarities, found for each name and on the Y-axis - the probability of a name to be invalid. The observed clusters are sparse and with possible poor quality, which as expected for the following reason - the number of fed on input names is not sufficient and they should be more. For the experiment 20 names, are used. They include the variations, creates a set of 200 inputs. The numbers of similarities from then on would be updated accordingly and the clusters should become denser. Nevertheless, this proves easily the original initial idea of the paper.

Now we can measure the overall confidence of the clustering approaches for **k**-s between two and twenty using (1). The results are shown in Figure 3:

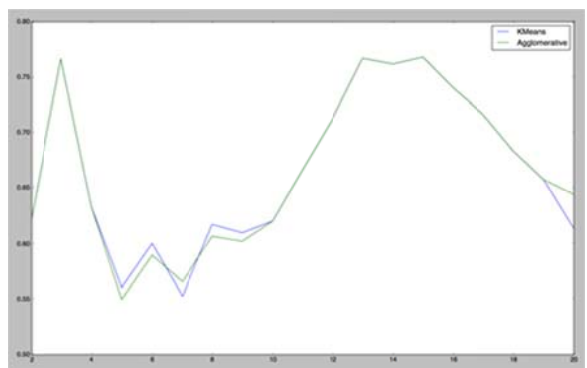

Figure 3

On the X-axis we observe the number of clusters and on the Y-axis is placed the respective evaluated Silhouette coefficient. As we expected the two approaches reach almost similar values of their confidence, although the **k-means** is obviously a bit more precise. Additionally, given the fact the computational effort for it is less than for the **agglomerative**, it is easy to conduct that for smaller number of clusters it would have a better overall efficiency. On the graph we also observe that for bigger number of clusters the trend is for **agglomerative** to dominate, which is again expected, because the smaller number of clusters, the greater number of recursive calls will be done.

## 5. Conclusion

In the present paper a new approach to reveal users' accounts is suggested. It combines a chaotic pseudo-random generator used for undesirable (false) account simulation, and cluster analysis for detection between false and real accounts. If there exist false accounts, then the actual ones would be used as a pattern and all neighbours will form clusters with the respective probability. This technique detects false accounts from the actual ones even when false and actual accounts generate with the same speed. On one hand, the advantage of the chaotic generator analysis there consists in the fact that it allows to determine the probability of false account appearance with name similar to this of some actual account. On the other hand, cluster analysis accelerates the process of forming the clusters of false accounts around each actual pattern.

One of important and significant application of the proposed in this paper security system is in on-line voting for balloting. Other application is in studying the social opinion by inquiries. Third application is associated with protection of the information in different user accounts of given system. This can help especially in the plagiarism detection.

The physical nature of the system is not important. It can be from different areas. First, as a system can consider separate technical device or aggregate, machine, phone, tablet, laptop or telecommunication system for transmitting the information or computer system for control some manufacturing process. Second, the system can be in the serving sphere – tourism (management of the hotels, visiting cities and famous places, book-keeping serving, on-line reserving the tickets. It can be generalized to any other personal data protection aerea.

In this sense, all their applications in different fields are pointed to increasing the data security in actual users' accounts. So for the applications it can lead to more secure users accounts for: e-banking, medical data exchange, health insurance, distance e-learning, on-line trading, on-line reserving hotels, on-line buying airways, railway and passage tickets, etc., since users can securely and freely benefit from modern tools for communication; as for voting, which could decrease the election costs.

The latter could result especially in higher voters' participation, to facilitate direct voters involvement in governmental decisions while preserving total privacy.

The present research can be developed in future works in the following directions: using the different linkage criterion for **agglomerative clustering**; applying different distance measuring functions and making automatic *k* detection.

## 7. Acknowledgements